\documentclass[12pt,a4paper]{article}
\usepackage[pdfstartview=FitH,colorlinks=true,linkcolor=black,anchorcolor=black,citecolor=black,urlcolor=black]{hyperref}
\usepackage[english]{babel}
\usepackage{amsmath,amssymb,titling,authblk}
\usepackage{slashed}
\usepackage{amsmath}
\usepackage{amscd}
\usepackage[normalem]{ulem}
\usepackage{appendix}
\usepackage{bbold}
\usepackage{bm}

\usepackage{pdflscape}
\usepackage{yfonts}
\usepackage{amscd}
\usepackage{epsfig}
 \usepackage{microtype}

\usepackage{cite}

\allowdisplaybreaks[4]
\numberwithin{equation}{section}

\usepackage{color}  

\definecolor{verde}{cmyk}{.83,.21,1,.08}
\definecolor{darkorchid}{rgb}{0.6, 0.2, 0.8}
\definecolor{rosso}{rgb}{1.0, 0.0, 0.0}

\def\ds{\stackrel{\star}{,}}
\def\nn{\nonumber}
\newcommand{\be}{\begin{equation}}
\newcommand{\ee}{\end{equation}}
\newcommand{\bea}{\begin{eqnarray}}
\newcommand{\eea}{\end{eqnarray}}
\newcommand{\ii}{\mathrm{i}}
\newcommand{\e}{\mathrm{e}}
\newcommand{\dd}{\mathrm{d}}
\newcommand{\del}{\partial}

\newcommand{\R}{\mathbb R}

\newcommand{\qed}{\nobreak \ifvmode \relax \else
      \ifdim\lastskip<1.5em \hskip-\lastskip
      \hskip1.5em plus0em minus0.5em \fi \nobreak
      \vrule height0.75em width0.5em depth0.25em\fi}

\begin{document}

\setlength{\droptitle}{-6pc}

\title{ Noncommutative  field theory  from angular twist\vspace{5pt}}

\renewcommand\Affilfont{\itshape}
\setlength{\affilsep}{1.5em}
\renewcommand\Authands{ and }

\author[1]{Marija Dimitrijevi\'c \'Ciri\'c\thanks{dmarija@ipb.ac.rs}}
\author[1]{Nikola Konjik\thanks{konjik@ipb.ac.rs}}
\author[2,3]{Maxim A.~Kurkov\thanks{max.kurkov@gmail.com}}
\author[2,3,4]{Fedele Lizzi\thanks{fedele.lizzi@na.infn.it}}
\author[2,3]{Patrizia Vitale\thanks{patrizia.vitale@na.infn.it}}
\affil[1]{
Faculty of Physics, University of Belgrade, Beograd, Serbia
\vspace{5pt}}
\affil[2]{Dipartimento di Fisica ``Ettore Pancini'', Universit\`{a} di Napoli {\sl Federico II}\vspace{5pt}, Napoli, Italy}
\affil[3]{INFN, Sezione di Napoli, Italy\vspace{5pt}}
\affil[4]{Departament de F\'{\i}sica Qu\`antica i Astrof\'{\i}sica and Institut de C\'{\i}encies del Cosmos (ICCUB),
Universitat de Barcelona. Barcelona, Spain}

\date{}

\maketitle 

\begin{abstract}\noindent
We consider a noncommutative field theory  with space-time  $\star$-commutators  based on an \emph{angular} noncommutativity, namely a solvable Lie algebra: the Euclidean in two dimension. The $\star$-product can be derived from a twist operator and it is shown to be invariant under twisted Poincar\'e transformations. In momentum space the noncommutativity manifests itself as a noncommutative  $\star$-deformed sum for the momenta, which allows for an equivalent definition of the $\star$-product in terms of  twisted convolution of plane waves. As an application, we analyze the $\lambda \phi^4$ field theory at one-loop and discuss its UV/IR behaviour. We also analyze the kinematics of  particle decay for two different situations: the first one corresponds to a splitting of space-time where only space is deformed, whereas the second one entails a non-trivial $\star$-multiplication for the time variable, while one of the three spatial coordinates stays commutative.
\end{abstract}
\newpage

\medskip

\section{Introduction}
The possibility that spacetime is described by a noncommutative geometry is a fascinating idea which has deep physical motivations going back to Bronstein~\cite{Bronstein} (for a more recent review see~\cite{WSS}). One way to study noncommutative spaces is to study field theories where the product is deformed into a $\star$-product so that the fields do not commute among themselves. The most studied field theory is the one described by the Gr\"onewold-Moyal product~\cite{Gronewold, Moyal} which adapts to spacetime the usual commutation rules of standard quantum mechanics~\cite{DFR}: 
\be
[x^\mu \stackrel{*}{,} x^\nu]=\ii\theta^{\mu\nu} \label{Moyalprod}
\ee 
with $\theta^{\mu\nu}$ a constant with the dimensions of length square. Field theories  on these spaces (for a review see~\cite{Szabo}) have a peculiar behaviour: some ultraviolet divergences are converted to infrared ones, a phenomenon known as UV/IR mixing~\cite{MinwallaSeibergVanRamsdong}. This is not only a characteristic of the Gr\"onewold-Moyal product, but in general of all translationally invariant products~\cite{GalluccioLizziVitale1, GalluccioLizziVitale2}. These theories are not Poincar\'e invariant, but they can be invariant under a twisted symmetry. It was shown in \cite{thetaPoincare,chaichian} that indeed, field theories based on the Gr\"onewold-Moyal $\star$-product are invariant under the twisted $\theta$-Poncar\'e symmetry. Another example of a noncommutative space-time invariant under a quantum symmetry is the $\kappa$-Minkowski space-time. It is an example of Lie algebra noncommutativity and it was first introduced~\cite{Lukierski1,Lukierski2} in the early 90's. Because of its symmetry properties, i.e.\ invariance under the $\kappa$-Poincar\'e Hopf algebra, $\kappa$-Minkowski represents an interesting playground for constructing physical models and investigating their behaviour under the noncommutative (NC) deformation.  Indeed, there is a whole family of  Lie-algebra based star-products, introduced in \cite{selene} (also see \cite{oriti} for a different approach); among those the one giving rise to the noncommutative space $\R^3_\lambda$ \cite{sheik}, which  has been widely studied \cite{kupr} in relation with quantum mechanics \cite{duflo}, quantum gravity \cite{oriti2} and quantum field theory \cite{vitalewallet, Gerevitalewallet, vitale, wallet, GereTuricWallet, JuricPoulainWallet}.  

In this paper we will discuss a scalar field theory for another Lie-algebra kind of noncommutative space which can be described by a Drinfel'd twist \cite{Drinfeld}. It is the space  characterized by the following commutation relations~\cite{KonjikDimitriejivicSamsarov} :
\bea
{}[x^3 \ds x^1]&=&-\ii\theta x^2 ,\nonumber\\
{}[x^3 \ds x^2]&=&\ii\theta x^1 \label{commrel}
\eea
all other commutators being zero, and with $\theta$ a constant with the dimension of length. The underlying Lie algebra can be recognized to be the Euclidean algebra $\mathfrak{e}(2)$, for which a star product realization   was already found in \cite{selene} by means of a Jordan-Schwinger map. One interesting property  of the  realization considered here relies on the fact that it is possible to exhibit a Drinfel'd twist for it.   The main motivation for this work, is that, although the commutation relations violate Poincar\'e symmetry, the space is the homogenous space of a \emph{twisted} Poincar\'e Hopf algebra. To this, let us define the following Drinfel'd twist:
\bea
\mathcal F(x,y) &=& \exp{\left\{-\frac{\ii\theta}{2}\left( \partial_{y^3}\left(x^2\partial_{x^1} - x^1 \partial_{x^2}\right)
-\partial_{x^3}\left(y^2\partial_{y^1} - y^1 \partial_{y^2}\right)
\right)\right\}}\nn\\
&=& \exp{\left\{\frac{\ii\theta}{2}\left( \partial_{y^3}\del_{\varphi_x}
-\partial_{x^3}\del_{\varphi_y}
\right)\right\}} \label{angtwist}
\eea
where for the last expression we have used cylindrical coordinates $x^1=\rho_x\cos\varphi_x$, $x^2=\rho_x\sin\varphi_x$, and an analogous expression for $y$. Since the  vector fields $\partial_{\varphi}$ and $\partial_3$ commute, the twist is an admissible one, because it satisfies the cocycle condition, a sufficient requirement for the associativity of the $\star$ product defined by:
\be
(f\star g)(x)=\mathcal F^{-1}(y,z) f(y) g(z)\bigg|_{x=y=z} = fg - \frac{\ii\theta}{2} (\del_\varphi f \del_3 g-\del_3 f\del_\varphi g) +O(\theta^2). \label{StarProduct}
\ee
Let us recall that the existence of a twist greatly simplifies  the construction of a differential calculus and the definition of noncommutative gauge and field theories~\cite{AschieriLizziVitale}.

This space is a variation of the previously mentioned $\kappa$-Minkowski space where the commutation relations are similarly of Lie-algebra type: $[x^0,x^i]=\ii\frac1\kappa x^i$, all other commutators vanishing. Field theory on the $\kappa$-Minkowski space-time is very much studied in the literature. In particular, scalar field theory and UV/IR mixing were discussed in \cite{KappaScalar1,KappaScalar2,KappaScalar3,KappaScalar4}, while NC gauge theory and coupling with fermions were introduced in \cite{WessGauge, DimitrijevicJonke}. Besides the already quoted \cite{selene}, where all three dimensional Lie algebras are analyzed, and related star products are introduced,  Lie algebra based NC spaces  and the corresponding field theories were studied also in~\cite{BertolamiGuisado} who considered nilpotent algebras (our case is solvable but not nilpotent) and~\cite{RobbinsSethi} which did not consider our noncommutative case. A common trait with the latter paper is the intertwining of UV/IR mixing with the violation of translational invariance. Field theories based on the $\mathfrak{su}(2)$ case as an instance of a simple algebra, also in connection with the UV/IR mixing, have been studied in \cite{vitalewallet,wallet}.

Relations~\eqref{commrel} can be substituted by an equivalent set with $x^3$ substituted by the time coordinate $x^0$. This was the choice  in the original paper~\cite{KonjikDimitriejivicSamsarov}. From an algebraic  point of view  there is no difference.   Conceptually  the two choices are clearly different: the former corresponds to a splitting of space-time where time remains commutative, while the latter singles out time as a noncommuting operator.  The computational difference appears as soon as one considers loop diagrams (see the remark in Sec.~\ref{scalar}).

The main result of the paper is the persistence  of the UV/IR mixing at one loop for an interacting scalar field theory, $\lambda \phi^{\star 4}$, which, although not retaining  Poincar\'e symmetry of its commutative analogue, is however invariant under twisted Poincar\'e transformations. 

The paper is organized as follows. In section \ref{ang} we introduce the noncommutative algebra under consideration and compute the $\star$-convolution of plane waves in terms of $\star$-sums of momenta.  We thus review the twisted Poincar\'e algebra introduced in \cite{KonjikDimitriejivicSamsarov} and relate our $\star$-sum to the deformed coproduct there defined. In section \ref{scalar} we introduce the $\lambda \phi^{\star 4}$ model and carefully study it in momentum space, at one loop. We find a deformed conservation of momenta which entails a deformation of the nonplanar correction to the propagator, with respect to the commutative case. We conclude that the phenomenon of UV/IR mixing, which was first discovered for Moyal noncommutativity, persists in the case under consideration.  

In order to better understand the phenomenon, in section \ref{threed} we pass to the $3$ dimensional case and repeat the analysis. We find that again, the planar diagram does not change with respect to the commutative case, while the nonplanar contribution exhibits mixing.

As an another application of our findings with respect to the deformed sum of momenta, we study in section \ref{particle} the kinematics of particles decays at the tree level. We conclude with a short summary and an appendix where the main calculations are performed. 

\section{Angular Noncommutativity}\label{ang}

The Abelian twist (\ref{angtwist}) is a special example of a more  general twist introduced in \cite{LukierskiAngTwist1,LukierskiAngTwist2,LukierskiAngTwist3}. The NC differential geometry induced by (\ref{angtwist}) was constructed in \cite{KonjikDimitriejivicSamsarov}, where also the NC field theory of a scalar field in the Reissner-Nordstr\" om background was investigated.

We remind   that the twist \eqref{angtwist} is  Abelian,  being based on  two commuting vector fields, $\del_{x^3}$ and $\del_\varphi$. Its form is reminiscent of the Moyal twist for the Moyal algebra $[x^1\stackrel\star,x^2]=\ii\theta$ , where the two vector fields are $\del_{x^1}$ and $\del_{x^2}$, but one should refrain from introducing the star commutator $[x^3\stackrel\star,\varphi]=\ii\theta$ since $\varphi$ is not a well defined continuous function. 
The following relations hold:
\bea
{}[x^3\stackrel\star,\rho]&=& 0  \nonumber\\
{}[x^3\stackrel\star,\e^{\ii\varphi}]&=& -\theta\e^{\ii\varphi} \nonumber\\
{}[x^3\stackrel\star, f(x^0,x^3,\rho,\varphi)]&=& \ii\theta\del_\varphi f  \label{commrelvarious}
\eea

For field theory it is useful to calculate the $\star$-product of two plane waves. This will enable the construction of the product in Fourier transform as some kind of twisted convolution. We remind that for the Gronew\"old-Moyal product~\eqref{Moyalprod} the analogous formula is $\e^{\ii p \cdot x}*\e^{\ii q \cdot x}=\e^{\ii (p+q)\cdot x + p_\mu\theta^{\mu\nu}q_\nu}$.  

While  the explicit calculations are  performed in the appendix, we reproduce here the relevant results. We have 
\be
e^{-\ii p \cdot x} \star e^{-\ii q\cdot  x} = e^{-\ii (p+_{\star}q)\cdot  x},  \label{planewaveprod}
\ee
where the $\star$-sum of the 4-momenta  is defined as follows:
\be
p+_{\star}q = R(q_3)p + R(-p_3)q, \label{starsum}
\ee
and $R$ is the following matrix:

\be
R(t) \equiv \left(
\begin{array}{cccc}
1 & 0 & 0 & 0 \\
0 & \cos{\left(\frac{\theta t}{2}\right)} & \sin{\left(\frac{\theta t}{2}\right)} & 0 \\
0 & -\sin{\left(\frac{\theta t}{2}\right)}  & \cos{\left(\frac{\theta t}{2}\right)} & 0 \\
0 & 0 & 0 & 1
\end{array}
\right) .\label{A}
\ee
The matrix $R$ corresponds to a rotation matrix in the $(p_1p_2)$ plane. The angle of  rotation is proportional to the noncommutativity parameter, and to the momenta involved. It reduces to the identity in the commutative limit $\theta \longrightarrow 0$ as well as in the low momentum limit. It is remarkable that, even though after the first two steps, \eqref{step1} and \eqref{step2}, one obtains expressions which are singular at $\frac{\theta q_3}{2} = \frac{\pi}{2} + \pi k$, $k\in Z$, the final expression 
\eqref{starsum} contains no singularities.

It is easy to see that the $\star$-sum  is noncommutative, but associative\footnote{This property reflects the associativity of the underlying $\star$-product.} and satisfies
\be
p+_{\star} (-p) = 0 \label{prop0}
\ee
for an arbitrary 4-vector $p$.
It is possible to generalize~\eqref{planewaveprod}
to the product of three plane waves, after extending  Eq. \eqref{starsum} to the sum of  three terms. We obtain 
\be
e^{-\ii p \cdot x} \star e^{-\ii q\cdot  x} \star e^{-\ii r\cdot  x}= e^{-\ii (p+_{\star}q+_\star r)\cdot  x},  
\ee
with\footnote{The deformation of the sum of momenta can be seen as a deformation of a Weyl system, an aspect which was already noted for $\kappa$-Minkowski in~\cite{Agostini}}
\be
p +_{\star} q +_{\star} r = R(r_3 + q_3)p + R(-p_3 + r_3)q  + R(-p_3 -q_3)r. \label{three}
\ee
By induction it is easily  shown that:
\be
p^{(1)} +_\star ... +_\star p^{(N)}  = \sum_{j=1}^N R\left(-\sum_{1\leq k<j}p_3^{(k)} + \sum_{j < k \leq N}p_3^{(k)}\right)p^{(j)}. \label{StarSumFull}
\ee
In the following subsection we will show that the $\star$-sum (\ref{StarSumFull}) can be related to the twisted coproduct of momenta $P_\mu$ in the twisted  Poincar\'e Hopf algebra, with angular twist defined in  (\ref{angtwist}).

\subsection{Twisted Poincar\'e Hopf algebra from the angular twist \label{TPHa}}

The full twisted Poincar\'e algebra and the corresponding differential calculus were introduced in \cite{KonjikDimitriejivicSamsarov}. Here we report the main results  that will be of importance for the subject of the paper, while referring to \cite{KonjikDimitriejivicSamsarov} for details, with the warning that $\partial_0$ is here  replaced by $\partial_3$.

 We will use the coordinate representation of Poincar\'e generators: \
 \bea
 P_\mu &=& -\ii\partial_\mu,  \nonumber \\
  M_{\mu\nu} &=&  \ii(\eta_{\mu\lambda}x^\lambda\partial_\nu - \eta_{\nu\lambda}x^\lambda\partial_\mu),
  \eea
  and $\eta_{\mu\nu}= (+,-,-,-)$. The Poincar\'e algebra is: 
\begin{eqnarray}
&& \lbrack P_{\mu },P_{\nu }] = 0,\quad \lbrack M_{\mu \nu },P_{\rho }]=\ii(\eta
_{\nu \rho }P_{\mu }-\eta _{\mu \rho }P_{\nu }),  \nn \\
&& \lbrack M_{\mu \nu },M_{\rho \sigma }] = \ii(\eta _{\mu \sigma }M_{\nu \rho
}+\eta _{\nu \rho }M_{\mu \sigma }-\eta _{\mu \rho }M_{\nu \sigma }-\eta
_{\nu \sigma }M_{\mu \rho }).  \label{TwistedPoincareAlg}
\end{eqnarray}
The twisted coproduct of momenta is given by
\begin{eqnarray}
&&\Delta^{\cal F} P_{0} = P_{0}\otimes 1 + 1\otimes P_{0},\nn\\
&&\Delta^{\cal F} P_{3} = P_{3}\otimes 1 + 1\otimes P_{3}, \label{TwistedCoproductMomenta} \\
&&\Delta^{\cal F} P_{1} = P_{1}\otimes \cos\left( \frac{\theta}{2}P_3 \right) + \cos\left(
\frac{\theta}{2}P_3 \right)\otimes P_{1} + P_{2}\otimes \sin\left( \frac{\theta}{2}P_3 \right) - \sin\left(
\frac{\theta}{2}P_3
\right)\otimes P_{2}, \nn\\
&&\Delta^{\cal F} P_{2} = P_{2}\otimes \cos\left( \frac{\theta}{2}P_3 \right) + \cos\left(
\frac{\theta}{2}P_3 \right)\otimes P_{2} - P_{1}\otimes \sin\left( \frac{\theta}{2}P_3 \right) + \sin\left(
\frac{\theta}{2}P_3
\right)\otimes P_{1}, \nn
\end{eqnarray}
while the twisted coproduct of Lorentz generators is:
\begin{eqnarray}
&&\Delta^{\cal F} M_{31} = M_{31}\otimes \cos\left( \frac{\theta}{2}P_3 \right) + \cos\left( \frac{\theta}{2}P_3
\right)\otimes M_{31} + M_{32}\otimes \sin\left( \frac{\theta}{2}P_3 \right) - \sin\left( \frac{\theta}{2}P_3
\right)\otimes M_{32}\nn\\
&& \hspace*{1.5cm} -P_1\otimes\frac{\theta}{2}M_{12}\cos\left( \frac{\theta}{2}P_3 \right) +
\frac{\theta}{2}M_{12}\cos\left( \frac{\theta}{2}P_3 \right)\otimes P_1 \nn\\
&& \hspace*{1.5cm} -P_2\otimes\frac{\theta}{2}M_{12}\sin\left( \frac{\theta}{2}P_3 \right) -
\frac{\theta}{2}M_{12}\sin\left( \frac{\theta}{2}P_3 \right)\otimes P_2,\nn\\
&&\Delta^{\cal F} M_{32} = M_{32}\otimes \cos\left( \frac{\theta}{2}P_3 \right) + \cos\left( \frac{\theta}{2}P_3
\right)\otimes M_{32} - M_{31}\otimes \sin\left( \frac{\theta}{2}P_3 \right) + \sin\left( \frac{\theta}{2}P_3
\right)\otimes M_{31}\nn\\
&& \hspace*{1.5cm} -P_2\otimes\frac{\theta}{2}M_{12}\cos\left( \frac{\theta}{2}P_3 \right) +
\frac{\theta}{2}M_{12}\cos\left( \frac{\theta}{2}P_3 \right)\otimes P_2\nn\\
&& \hspace*{1.5cm} +P_1\otimes\frac{\theta}{2}M_{12}\sin\left( \frac{\theta}{2}P_3 \right) +
\frac{\theta}{2}M_{12}\sin\left( \frac{\theta}{2}P_3 \right)\otimes P_1,\nn\\
&&\Delta^{\cal F} M_{30}=M_{30}\otimes 1 + 1\otimes M_{30} -\frac{\theta}{2}P_0\otimes M_{12}
+\frac{\theta}{2}M_{12}\otimes P_0 ,\label{TwistedCoproductLorGen}
\end{eqnarray}
\begin{eqnarray}
&&\Delta^{\cal F} M_{12}=M_{12}\otimes 1+1\otimes M_{12},\nn\\
&&\Delta^{\cal F} M_{10} = M_{10}\otimes \cos\left( \frac{\theta}{2}P_3 \right) + \cos\left( \frac{\theta}{2}P_3
\right)\otimes M_{10} + M_{20}\otimes \sin\left( \frac{\theta}{2}P_3 \right) - \sin\left( \frac{\theta}{2}P_3
\right)\otimes M_{20}\nn\\
&&\Delta^{\cal F} M_{20} = M_{20}\otimes \cos\left( \frac{\theta}{2}P_3 \right) + \cos\left( \frac{\theta}{2}P_3
\right)\otimes M_{20} - M_{10}\otimes \sin\left( \frac{\theta}{2}P_3 \right) + \sin\left( \frac{\theta}{2}P_3
\right)\otimes M_{10} .\nn
\end{eqnarray}
The coproducts of momenta $P_0$ and $P_3$ and  of $M_{12}$, the generator of the rotation in the $x^1x^2$ plane, remain undeformed (primitive). All other coproducts are clearly deformed. Before we introduce the corresponding differential calculus and the integration, we comment on the twisted coproduct of momenta and its relation with the $\star$-sum of momenta (\ref{StarSumFull}).

Let us suppose that a field  $\phi_p$ is an eigenvector of the momentum operator $P_\mu$ with the eigenvalue $p_\mu$:
\begin{equation}
P_\mu \phi_p = p_\mu \phi_p. \nn  
\end{equation}
Plane waves $e^{-ip\cdot x}$ are one example of these states. In order to check if the field $\phi_p\star \phi_q$ is again an eigenstate of the momentum operator $P_\mu$ we have to calculate:
\begin{equation}
P_\mu (\phi_p\star \phi_q) = \mu_\star \{ \Delta^{\cal F}P_\mu (\phi_p\otimes \phi_q) \}  ,\label{StSumCoproduct1}
\end{equation}
where $\mu_\star$ represents  the $\star$-product introduced in (\ref{StarProduct}):
\be
 \mu_\star\left(\phi_p\otimes \phi_q\right) = \phi_p\star \phi_q.
 \ee
In components, using (\ref{TwistedCoproductMomenta}), we obtain:
\begin{eqnarray}
P_0 \left(\phi_p\star \phi_q\right) &=& (p+q)_0 \left(\phi_p\star \phi_q\right) = (p+_\star q)_0 (\phi_p\star \phi_q), \nn\\
P_3 (\phi_p\star \phi_q) &=& (p+q)_3 (\phi_p\star \phi_q)= (p+_\star q)_3 (\phi_p\star \phi_q) ,\nn\\
P_1 (\phi_p\star \phi_q) &=& \mu_\star\Big\{\Big(P_{1}\otimes \cos\left( \frac{\theta}{2}P_3 \right) + \cos\left(
\frac{\theta}{2}P_3 \right)\otimes P_{1} + P_{2}\otimes \sin\left( \frac{\theta}{2}P_3 \right) \nn\\
&&- \sin\left(
\frac{\theta}{2}P_3\right)\otimes P_2\Big) \big(\phi_p\otimes\phi_q\big)\Big\}\nonumber\\ 
&=& \left(p_{1} \cos\left( \frac{\theta}{2}q_3 \right) + q_1\cos\left(
\frac{\theta}{2}p_3 \right)  + p_{2} \sin\left( \frac{\theta}{2}q_3 \right) - q_2\sin\left(
\frac{\theta}{2}p_3 \right)\right)\nn\\
&&(\phi_p\star \phi_q) = (p+_\star q)_1 (\phi_p\star \phi_q), \label{StSumCoproduct2}\\
P_2 (\phi_p\star \phi_q) &=& \mu_\star\Big\{\Big(P_{2}\otimes \cos\left( \frac{\theta}{2}P_3 \right) + \cos\left(
\frac{\theta}{2}P_3 \right)\otimes P_{2} - P_{1}\otimes \sin\left( \frac{\theta}{2}P_3 \right) \nn\\
&& + \sin\left(
\frac{\theta}{2}P_3
\right)\otimes P_{1}\Big)\big(\phi_p\otimes\phi_q\big)\Big\}\nn\\ 
&=& \left(p_{2} \cos\left( \frac{\theta}{2}q_3 \right) + q_{2}\cos\left(
\frac{\theta}{2}p_3 \right)  - p_{1} \sin\left( \frac{\theta}{2}q_3 \right) + q_{1}\sin\left(
\frac{\theta}{2}p_3 \right)\right)\nn\\
&&(\phi_p\star \phi_q) = (p+_\star q)_2 (\phi_p\star \phi_q). \nn
\end{eqnarray} 
Thus, we see that the field $\phi_p\star \phi_q$ is indeed an eigenvector of the momentum operator $P_\mu$ and its eigenvalue is given by the $\star$-sum (\ref{starsum}).

We can repeat the calculation for the case of the $\star$-product of three fields $\phi_p\star\phi_q\star\phi_r$. We just sketch the calculation for the $P_1$ component, as the other components can be calculated following the same steps. We have
\begin{eqnarray}
P_1 \left(\phi_p\star \phi_q\star \phi_r\right) &=&  \mu_\star \left 
\{ \Delta^{\cal F}P_1 (\phi_p\otimes \phi_q\otimes \phi_r) \right\} \nn\\
&=&  \mu_\star \left\{ \left( P_{1}\otimes \cos\left( \frac{\theta}{2}P_3 \right) + \cos\left(
\frac{\theta}{2}P_3 \right)\otimes P_{1} + P_{2}\otimes \sin\left( \frac{\theta}{2}P_3 \right)\right. \right. \nn\\
&&\left.\left. - \sin\left(
\frac{\theta}{2}P_3
\right)\otimes P_{2}\right)\left(\phi_p\otimes (\phi_q\otimes \phi_r)\right) \right\} \nn\\
&=&  \mu_\star \left\{ \left(  P_{1}\phi_p \otimes \cos\left( \frac{\theta}{2}P_3 \right)(\phi_q\otimes \phi_r) + \cos\left(
\frac{\theta}{2}P_3 \right)\phi_p\otimes P_{1}(\phi_q\otimes \phi_r)\right.\right. \nn\\
&&\left.\left.+ P_{2}\phi_p\otimes \sin\left( \frac{\theta}{2}P_3 \right)(\phi_q\otimes \phi_r) - \sin\left(
\frac{\theta}{2}P_3
\right)\phi_p\otimes P_{2} (\phi_q\otimes \phi_r )\right) \right\} \nn\\
&=& \left( p_{1} \cos\left( \frac{\theta}{2}(q+r)_3 \right)  + p_{2} \sin\left( \frac{\theta}{2}(q+r)_3 \right)\right.\nn\\ 
&&+ q_{1} \cos\left( \frac{\theta}{2}(-p+r)_3 \right)  + q_{2} \sin\left( \frac{\theta}{2}(-p+r)_3 \right)\nn\\
&& \left.r_{1} \cos\left( \frac{\theta}{2}(-p-q)_3 \right)  + r_{2} \sin\left( \frac{\theta}{2}(-p-q)_3 \right) \right) (\phi_p\star \phi_q\star \phi_r) \nn\\
&=& (p+_\star q +_\star r)_1 (\phi_p\star \phi_q\star \phi_r) \label{StSumCoproduct3}
\end{eqnarray} 
where we have used:
\begin{itemize}
\item the coassociativity of the coproduct (\ref{TwistedCoproductMomenta}) in the second line;
\item the coproduct of momenta (\ref{TwistedCoproductMomenta}) and in addition the following coproducts
\begin{eqnarray}
\Delta^{\cal F} \left( \cos\left( \frac{\theta}{2}P_3 \right) \right) &=& \cos\left( \frac{\theta}{2}P_3 \right)\otimes \cos\left( \frac{\theta}{2}P_3 \right) - \sin\left( \frac{\theta}{2}P_3 \right)\otimes \sin\left( \frac{\theta}{2}P_3 \right) ,\nn\\
\Delta^{\cal F} \left( \sin\left( \frac{\theta}{2}P_3 \right) \right) &=& \cos\left( \frac{\theta}{2}P_3 \right)\otimes \sin\left( \frac{\theta}{2}P_3 \right) + \sin\left( \frac{\theta}{2}P_3 \right)\otimes \cos\left( \frac{\theta}{2}P_3 \right) .\nn
\end{eqnarray}

\end{itemize}
Again we conclude that a field $\phi_p\star \phi_q\star \phi_r$ is an eigenvector of the momentum operator $P_\mu$ and its eigenvalue is given by the $\star$-sum $(p+_\star q +_\star r)_\mu$.

The calculation can be extended to the $\star$-product of $n$ fields with the same result.  Summarizing, the twist of the Poincar\'e   symmetry manifests itself in the fact that the coproducts are modified, so that the action of generators on multiparticle states is deformed, while the  algebra itself remains undeformed. An interesting property  of the angular twist (\ref{angtwist}) is that the ensuing deformation is easily expressed in terms of $\star$-sums of one-particle momenta.

Let us proceed to introducing the basic elements of the corresponding differential calculus. The wedge product of two forms of arbitrary degree, $\omega_1$ and $\omega_2$, is deformed into the $\star$-wedge product:
\begin{equation}
(\omega_1 \wedge_\star \omega_2)(x) =\mathcal F^{-1}(y,z) \omega_1(y)\wedge \omega_2 (z)\bigg|_{x=y=z}. \label{WedgeStar}
\end{equation}
This $\star $-product is graded noncommutative and associative. The usual (commutative) exterior derivative satisfies: 
\begin{eqnarray}
\mathrm{d} (f\star g) &=& \mathrm{d}f\star g + f\star \mathrm{d}g,  \notag \\
\mathrm{d}^2 &=& 0.  \label{Differential}
\end{eqnarray}
The first property if fulfilled because the usual exterior derivative
commutes with Lie derivatives that enter in the definition of the $\star
$-product. Therefore, we will use the usual exterior derivative as the
noncommutative exterior derivative.

Since the twist (\ref{angtwist}) is Abelian, the cyclicity of the integral holds
\begin{equation}
\int \omega _{1}\wedge _{\star }\dots \wedge _{\star }\omega
_{p}=(-1)^{d_{1}\cdot d_{2}\dots \cdot d_{p}}\int \omega _{p}\wedge _{\star
}\omega _{1}\wedge _{\star }\dots \wedge _{\star }\omega _{p-1}, 
\end{equation}
with $d_{1}+d_{2}+\dots +d_{p}=4$. It can be shown that the twist (\ref{angtwist}) fulfils an even
stronger requirement. Namely, \cite{pachol
}
one can check that the  $\star$-products of functions is indeed closed
\begin{equation}
\int {\rm d}^ 4x\, f\star g  = \int {\rm d}^ 4x\, g\star f = \int {\rm d}^ 4x\, f\cdot g .\label{IntCyclfg}
\end{equation}
The last property in general does not hold for coordinate dependent $\star$-products, as for example  $\kappa$-Minkowski $\star$products \cite{KappaScalar1,KappaScalar2, KappaScalar3,KappaScalar4,DimitrijevicJonke, pachol}, or $\mathfrak{su}(2)$ ones \cite{kupr} .

\section{Scalar Field Theory}\label{scalar}
Consider a scalar theory on our noncommutative space described by

\be
S[\phi] = \int_{\mathbb{R}^4} \dd^4x\, \left(\frac{1}{2}\partial_{\mu}\phi(x)\star\partial^{\mu}\phi(x) - \frac{1}{2} m^2 \phi(x)\star\phi(x) - \frac{\lambda}{4!}\phi(x)^{\star 4}\right). \label{action}
\ee

Because of the closure of the $\star$-product, stated in Eq.~\eqref{IntCyclfg} it is possible to replace the $\star$-product in all  quadratic terms by the usual (pointwise) one. Nevertheless, we will keep it for the forthcoming discussion for clarity. A consequence of this fact is that the free propagators are the same as in the commutative theory. The interacting one instead will change, as we will see with a one-loop calculation.

\subsection{Deformed Conservation of Momentum}
Expanding the field $\phi(x)$ in its Fourier modes 
\be
\phi(x) = \frac{1}{(2\pi)^2}\int_{\mathbb{R}^4} \dd^4p \, e^{-\ii px} \widetilde\phi(p)
\ee
one arrives at the following expression for the classical action in  momentum space
\bea
S[\phi] &=& \int_{\mathbb{R}^4\times \mathbb{R}^4} \dd p \,\dd q\,\, \frac{1}{2}\left(-p_{\mu}q^{\mu}\widetilde\phi(p) \widetilde\phi(q) - m^2 \widetilde\phi(p)\widetilde\phi(q)\right)\delta^{(4)}\left(p +_\star q\right) \label{paction} \\
&-&\frac{1}{(2\pi)^4}\frac{\lambda}{4!} \int_{\left(\mathbb{R}^4\right)^{\times 4}} \dd p\, \dd q\,\dd r\,\dd s\,\, \widetilde\phi(p)\widetilde\phi(q)\widetilde\phi(r)\widetilde\phi(s)
\delta^{(4)}\left(p +_\star q+_\star r+_\star s\right) \nonumber.
\eea 
One can see that \emph{the only} difference of this expression with respect to the commutative case is the presence of the $\star$-sum instead of the usual sum in the delta functions. These $\delta$ functions encode  the conservation of momentum in the corresponding vertices. Therefore, the  main difference with respect to the commutative case is the twisted conservation of momentum. If just two momenta are involved in the game (c.f.\ the quadratic terms of the action) the deformed conservation of the momenta is equivalent to the usual one, the rotation in the $x^1x^2$ plane does not matter. Because of the  block diagonal structure of the matrix \eqref{A} the conservation of the zeroth and third components of the momenta are not affected by the deformation. Then the  property becomes obvious:
\be
\delta^{(4)}\left(p+_\star q\right) = \delta^{(4)}(R(q_3)p + R(-p_3)q) = \delta^{(4)}(R(q_3)(p + q))  = \delta^{(4)}(p + q), \label{prop1}
\ee
where we used the fact that, for any non degenerate $n$ by $n$ matrix $M$ and $n$-component (vector-valued) function $V$,  the $n$-dimensional delta function satisfies the identity
\be
\delta^{(n)}\left(MV\right) = \frac{1}{|\det{M}|}\,\delta^{(n)}(V), \label{deltaprop}
\ee 
and the fact that $\det{R} = 1$.
The property \eqref{prop1} and the associativity of the $\star$-sum  immediately imply a slightly more general relation
\be
\delta^{(4)}\left(p^{(1)} +_{\star}... +_{\star}p^{(k)}+_{\star}... +_{\star}p^{(N)}\right) = \delta^{(4)}\left(p^{(1)} +_{\star}... +p^{(k)}+_{\star}... +_{\star}p^{(N)}\right), \label{prop2}
\ee
i.e.\ upon the sign of the delta function one of the $\star$-sums can be replaced by the usual one. This property is in agreement with the (known) fact that in each term  of the action \eqref{action} one of the
star products  can be replaced by the standard pointwise one.

\subsection{One-loop propagator and IR/UV mixing}

\begin{figure}[htbp]
\epsfxsize=4.5 in
\bigskip
\centerline{\epsffile{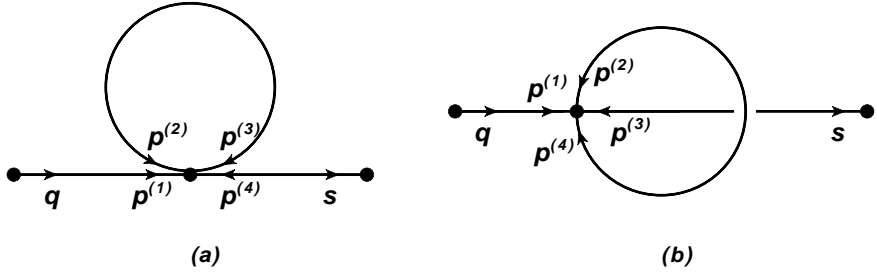}}
\caption{\sl The planar (a) and nonplanar (b) one loop correction
to the propagator, which correspond to \eqref{planar} and \eqref{nonplanar} respectively. \label{diagram}}
\end{figure}

To study the UV/IR behaviour of the model,  we will consider the one loop correction to the free propagator. In a noncommutative theory there are two of these diagrams, planar and nonplanar, shown in Fig.~\ref{diagram}.

The planar and non planar contributions to the propagator, which are of the order of $\lambda$, are proportional to 
\bea \!\!\!\!\!\!\!\!\!\!\!\!\!\!\!\!\!\!
\mbox{Pl.}&\propto&\frac{1}{ m^2 - q^2 - \ii 0^{+}}\cdot\frac{1}{m^2 -s^2 - \ii 0^{+}}\cdot \int_{\left(R^{4}\right)^{\times 4}} dp^{(1)}dp^{(2)}dp^{(3)}dp^{(4)}\,\delta^{(4)}\left(p^{(2)}+p^{(3)}\right)\, \times \nonumber \\
&\times&\delta^{(4)}\left(q - p^{(1)}\right)\,\delta^{(4)}\left(p^{(4)} + s\right)\,\delta^{(4)}\left(p^{(1)} +_{\star}p^{(2)}+_{\star}p^{(3)}+_{\star}p^{(4)}\right) \cdot\frac{1}{  m^2 - \left(p^{(2)}\right)^2 - \ii 0^{+}}  \label{planar}
\eea 
and 
\bea  \!\!\!\!\!\!\!\!\!\!\!\!\!\!\!\!\!\!
\mbox{NPl.}&\propto& \frac{1}{ m^2 - q^2 - \ii 0^{+}}\cdot\frac{1}{m^2 -s^2 - \ii 0^{+}}  \cdot \int_{\left(R^{4}\right)^{\times 4}} dp^{(1)}dp^{(2)}dp^{(3)}dp^{(4)}\,\delta^{(4)}\left(p^{(2)} + p^{(4)}\right)\, \times  \nonumber\\
&\times&\delta^{(4)}\left(q - p^{(1)}\right)\,\delta^{(4)}\left(p^{(3)} + s\right)\,\delta^{(4)}\left(p^{(1)} +_{\star}p^{(2)}+_{\star}p^{(3)}+_{\star}p^{(4)}\right)\cdot\frac{1}{  m^2 - \left(p^{(2)}\right)^2 - \ii 0^{+}}  \label{nonplanar}
\eea 
respectively. Hereafter $s^2 = s_{\mu}s_{\nu}\eta^{\mu\nu}$, $q^2 = q_{\mu}q_{\nu}\eta^{\mu\nu}$ and etc. Since we are only interested in discussing the structure of divergences, in order to simplify the notation, we have omitted to indicate constants such as  symmetry factors, the parameter $\lambda$ and powers of $2\pi$. 

Whilst the delta functions, which involve the star sums, are identical, the remaining parts of the integrands map onto each other upon the replacement 
$p^{(3)} \leftrightarrow p^{(4)}$. Since the $\star$-sum is not a commutative operation, the delta functions are not invariant upon such a replacement, therefore the two contributions are actually different.

The former is identical to the corresponding expression in the commutative case. Indeed the delta function which involves all the $\star$-sums appears together with $\delta^{(4)}\left(p^{(2)}+p^{(3)}\right)$, and 
\bea
&&\delta^{(4)}\left(p^{(2)}+p^{(3)}\right)\delta^{(4)}\left(p^{(1)} +_{\star}p^{(2)}+_{\star}p^{(3)}+_{\star}p^{(4)}\right)\nonumber \\
&&= \delta^{(4)}\left(p^{(2)}+p^{(3)}\right)\delta^{(4)}\left(p^{(1)} +_{\star}\left(p^{(2)}+_{\star}(-p^{(2)})\right)+_{\star}p^{(4)}\right) \nonumber\\
&& = \delta^{(4)}\left(p^{(2)}+p^{(3)}\right)\delta^{(4)}\left(p^{(1)} +_{\star} p^{(4)}\right)  \nonumber\\
&& = \delta^{(4)}\left(p^{(2)}+p^{(3)}\right)\delta^{(4)}\left(p^{(1)} + p^{(4)}\right), \label{calc}
\eea
where we used the associativity and the properties \eqref{prop0} and \eqref{prop1} of the $\star$-sum. We emphasize that the last line of the calculation \eqref{calc} does not contain the $\star$-sum, i.e.\ the same expression holds in the commutative case.
Substituting \eqref{calc} in \eqref{planar} we immediately obtain 
\be
\mbox{Pl.} \propto \frac{1}{ m^2 - q^2 - \ii 0^{+}}\cdot\frac{1}{m^2 -s^2 - \ii 0^{+}}\cdot \delta^{(4)}(q-s) \int_{\mathbb{R}^4} d p\,\, \frac{1}{ m^2 - p^2 - \ii 0^{+} }. \label{planar1}
\ee 
First we perform the Wick rotation $p_0 \longrightarrow \ii p_0$ of the integration path over the energy $p_0$. After that we regularize the divergent integral introducing an UV cutoff $\Lambda$ and integrate over the momenta $p$ such that\footnote{Hereafter $|p|_{\mathrm{Eucl}}$ stands for  $\sqrt{p_{\mu}p_{\nu} \delta^{\mu\nu} }$.} $|p|_{\mathrm{Eucl}} \leq \Lambda$. Then, passing to  spherical coordinates, we immediately find that  the remaining integral in \eqref{planar1} equals to 
\be
 2\pi^2\int_0^{\Lambda}d|p|_{\mathrm{Eucl}}\, \frac{|p|_{\mathrm{Eucl}}^3}{|p|_{\mathrm{Eucl}}^2 + m^2} = \pi^2\Lambda^2 - \pi^2 m^2\log{\frac{\Lambda^2 + m^2}{m^2}} = 
 \pi^2\Lambda^2  -\pi^2 m^2\log{\frac{\Lambda}{\mu}} + ...,
\ee
where ``..." stands for the UV finite part and the arbitrary dimensionful parameter $\mu$ describes the ambiguity of the subtraction of the logarithmic divergence\footnote{In commutative theories this parameter is the normalization point and its presence gives rise to the renormalization group flow, whose discussion goes beyond the scope of this paper.}. Collecting all together we arrive at the following answer for the planar contribution 
\be
\mathrm{Pl.} = \frac{\ii}{ m^2 - q^2 - \ii 0^{+}}\cdot\frac{1}{m^2 -s^2 - \ii 0^{+}} \cdot \delta^{(4)}(q-s)\pi^2\left(  \Lambda^2  - m^2\log{\frac{\Lambda}{\mu}} 
\right) + \mbox{UV finite terms},
\ee 
which contains the well known quadratic and logarithmic divergences.

Now let us explore the non planar expression \eqref{nonplanar}. One can easily carry out some of the integrations thanks to the presence of the first three delta functions:
\bea
\mbox{NPl.}&=&\frac{1}{ m^2 - q^2 - \ii 0^{+}}\cdot\frac{1}{m^2 -s^2 - \ii 0^{+}}\, \times \nonumber\\
&\times&\int_{R^{4} } dp\,\delta^{(4)}\left( q+_{\star}(-p)+_{\star}(-s) +_{\star} p)\right)\,\frac{1}{ m^2 - p^2 - \ii 0^{+}} \label{nonplanar1},
\eea 
where we replaced $-p^{(2)}$ by $p$. Let us take a closer look at the delta function in \eqref{nonplanar1}.
Substituting $r = -p$ in \eqref{three} and using the property \eqref{deltaprop} of the delta function with $M = R$ we immediately obtain
\be
\delta^{(4)}\left( q+_{\star}(-p)+_{\star}(-s) +_{\star} p)\right) =\delta^{(4)}\left(\left(R(q_3)- R(-q_3)\right)p + R(-p_0)q
-  R(p_0)s
\right).  \label{deltainterm0}
\ee
The situation qualitatively differs from  the planar case: the momenta $p$ is under the sign of the delta function, therefore the number of the (one dimensional) integrations in the loop is actually less than four.  The matrix, which acts on $p$ reads:
\be
R(q_3)- R(-q_3) =\left[ \begin {array}{cccc} 0&0&0&0\\ \noalign{\medskip}0&0&2\,\sin \left( \frac{\theta q_3}{2} \right) &0\\ \noalign{\medskip}0&-2\,\sin \left( \frac{\theta q_3}{2} \right) &0&0\\ \noalign{\medskip}0&0&0&0
\end {array} \right] \label{matrix}
\ee
For the further steps it is convenient to parametrise
\bea
q_1 &=& |q| \cos{\psi}, \quad q_2 = |q| \sin{\psi} \nonumber\\
s_1 &=& |s| \cos{\zeta}, \quad s_2 = |s| \sin{\zeta}.
\eea
In order to avoid confusions we emphasize that $|q|$ and $|s|$ stand for the lengths of the \emph{projections} of $q$ and $s$ onto $x^1x^2$-plane.
Hereafter we suppose that both the noncommutativity parameter $\theta$ and the component $q_3$ of the external momenta $q$ are different from zero (or an integer multiple of $2\pi$).
Direct calculation shows that the delta function \eqref{deltainterm0} equals to
\be
\frac{1}{4\, \left( \sin \left( \frac{\theta q_3}{2} \right)  \right) ^{2}} \,\,\delta(q_3 - s_3)\, \delta(q_0 - s_0)\delta\left(p_1 - \mathcal{P}_1(q,s,p_3)\right)\delta
\left(p_2 - \mathcal{P}_2(q,s,p_3)\right), \label{deltainterm}
\ee
where we applied the rule \eqref{deltaprop} to the two-dimensional delta function in the $x^1x^2$-plane, in particular 
the overall factor is nothing but the inverse of the determinant of the $2\times 2$ ``central" block of the matrix \eqref{matrix}. The functions $\mathcal{P}_1(q,s,p_3)$ and
 $\mathcal{P}_2(q,s,p_3)$, 
which are given by
\bea
\mathcal{P}_1(q,s,p_3) &=& \,{\frac {-|s|\sin \left( - \frac{\theta p_3}{2}+\zeta \right) +|q|\sin
 \left(  \frac{\theta p_3}{2}+\psi \right) }{2\sin \left(  \frac{\theta q_3}{2} \right) }} ,\nonumber\\
\mathcal{P}_2(q,s,p_3) &=&{\frac {|s|\cos \left( - \frac{\theta p_3}{2}+\zeta \right) -|q|\cos
 \left(  \frac{\theta p_3}{2}+\psi \right) }{2\sin \left(  \frac{\theta q_3}{2}\right) }},
\eea
are the first and the second component of the solution $p_*$ of the algebraic system 
\be
\left\{\begin{array}{c}
\left(\left(R(q_3)- R(-q_3)\right)p_* + R(-p_3)q
-  R(p_3)s\right)_1 = 0 \\
\left(\left(R(q_3)- R(-q_3)\right)p_* + R(-p_3)q
-  R(p_3)s\right)_2 = 0
\end{array}
\right. ,
\ee
where, the subscripts 1 and 2 label the 4-vector components.

So we see that the key difference with respect to the commutative case is the opportunity to integrate out two out of four components of the momenta $p$ using the last two delta functions in \eqref{deltainterm}. This implies that the UV divergence can be at most logarithmic but not quadratic. The integral over momenta in \eqref{nonplanar1} reads: 
\bea
&&\int_{R^{4} } dp\,\delta^{(4)}\left(p +_{\star} q+_{\star}(-p)+(-s))\right)\,\frac{1}{ m^2 - p^2 - \ii 0^{+}} = \frac{1}{4\, \left( \sin \left( \frac{\theta q_3}{2} \right)  \right) ^{2}} \,\,\delta(q_0 - s_0)\, \delta(q_3 - s_3) \times\nonumber \\
&&
\times\int_{R^2} dp_0\,dp_3 \, \,\,\frac{1}{- p_0^2 + p_3^2+\left(\mathcal{P}_1(q,s,p_3)\right)^2 + \left(\mathcal{P}_2(q,s,p_3)\right)^2 +m^2 - \ii 0^{+}}. \label{interm2beta}
\eea
Performing the Wick rotation $p_0 \longrightarrow \ii p_0$ and introducing 
the ultraviolet cutoff $\Lambda$ we regularize the integral over $p_0$ and $p_3$ restricting the area of integration to $p_1^2 + p_3^2 <\Lambda^2$.
\bea
&& \frac{\ii}{4\, \left( \sin \left( \frac{\theta q_3}{2} \right)  \right) ^{2}} \,\,\delta(q_0 - s_0)\, \delta(q_3 - s_3) \times\nonumber \\
&&
\times\int_{p_0^2 +p_3^2 < \Lambda^2} dp_0\,dp_3 \, \,\,\frac{1}{ p_0^2 + p_3^2+\left(\mathcal{P}_1(q,s,p_3)\right)^2 + \left(\mathcal{P}_2(q,s,p_3)\right)^2 +m^2 }. \label{interm2}
\eea 
At this point an important remark is in order. 
Since
\be
\left(\mathcal{P}_1(q,s,p_3)\right)^2 + \left(\mathcal{P}_2(q,s,p_3)\right)^2 = {\frac {2\,|s|\,|q| \left( 1-\cos \left( -\theta\,p_{{3}}+\zeta-\psi
 \right)  \right) + \left( |q|-|s| \right) ^{2}}{ 4\left( \sin \left(\frac{\theta q_3}{2} \right)  \right) ^{2}}} \label{p1p2}  
\ee 
the following estimate holds
\be
{\frac { \left( |q|-|s| \right) ^{2}}{ 4\left( \sin \left(\frac{\theta q_3}{2} \right)  \right) ^{2}}} \leq  \left(\mathcal{P}_1(q,s,p_3)\right)^2 + \left(\mathcal{P}_2(q,s,p_3)\right)^2 \leq {\frac { \left( |q|+|s| \right) ^{2}}{ 4\left( \sin \left(\frac{\theta q_3}{2} \right)  \right) ^{2}}}. \label{estimate}
\ee

\vspace{0.9cm}

{\small{{\noindent\emph{ {\bf Remark.} Let us comment what would happen if we considered noncommutative time $x_0$ instead of $x_3$ in the commutation relations \eqref{commrelvarious}.
In this situation the Fourier conjugated  momenta $q_0$ would enter in the combination $\mathcal{P}_1^2+\mathcal{P}_2^2$ instead of $p_3$ in \eqref{interm2} and\eqref{p1p2}. The Wick rotation  $p_0 \longrightarrow \ii p_0$ will not help to compute the nonplannar contribution.
Indeed, setting external momenta $q$ and $s$ such that the angles $\psi$ and $\zeta$ are equal, one finds that after the Wick rotation the integrand of the nonplannar contribution equals to:
 \be
 \frac{1}{m^2 + p_0^2 +\left(\mathcal{P}_1(q,s,\ii p_0)\right)^2 + \left(\mathcal{P}_2(q,s,\ii p_0)\right)^2} = \frac{1}{m^2 +p_0^2 +\frac {2\,|s|\,|q| \left( 1-\cosh \left( \theta\,p_{{0}}
 \right)  \right) + \left( |q|-|s| \right) ^{2}}{ 4\left( \sin \left(\frac{\theta q_3}{2} \right)  \right) ^{2}}}. \label{discu}
 \ee
The denominator of this expression is obviously negative at large enough $p_0$ but positive, when $p_0$ is small, hence   \eqref{discu} exhibits a singularity at finite $p_0$, which prevents integration over $p_0$.}} } } 

The right hand side of the inequality \eqref{estimate} does not depend on the momenta $p$, therefore the presence of  \eqref{p1p2} in the denominator of the integrand in the righthand side of \eqref{interm2} does not affect the UV asymptotics of the integrand, which can be presented as follows: at $p_0^2 +p_3^2 \rightarrow \infty$
\be
\frac{1}{p_0^2 + p_3^2+\left(\mathcal{P}_1(q,s,p_3)\right)^2 + \left(\mathcal{P}_2(q,s,p_3)\right)^2 +m^2} \simeq 
\frac{1}{p_0^2 + p_3^2 +\mu^2} + \mathcal{O}\left(\frac{1}{\left(p_0^2 +p_3^2\right)^2}\right), \label{asymp}
\ee
where $\mu$ stands for an arbitrary dimensionful parameter. Its role will be clear later on.

At $m\neq0$ the integral \eqref{interm2} is obviously free of infrared divergences. Therefore we should analyse only the ultraviolet limit.   
Note that the contribution of the terms in \eqref{asymp} which are of the order of $\mathcal{O}\left(\frac{1}{\left(p_0^2 +p_3^2\right)^2}\right)$ is UV finite,
and we obtain:
\be
\mbox{NPl.} = \frac{\ii}{ m^2 - q^2 - \ii 0^{+}}\,\frac{1}{m^2 - s^2 - \ii 0^{+}}\, \delta(q_0 - s_0)\, \delta(q_3 - s_3)\,\frac{\pi^2}{2\, \left( \sin \left( \frac{\theta q_3}{2} \right)  \right) ^{2}} \,\ln{\left(\frac{\Lambda}{\mu}\right)} + \mbox{finite part}. \label{nonplanans}
\ee 
Let us  comment about the parameter $\mu$. Its computational role is quite clear:  it ensures the IR convergence of the  leading term of the UV asymptotics~\eqref{asymp} in the integral~\eqref{interm2}. The fact that the UV divergent  part depends on the arbitrary parameter $\mu$ describes the ambiguity of the subtraction of the logarithmic divergence, which is present already in the commutative case.  

Let us summarize our findings:
\begin{itemize}
\item The planar contribution is not affected by noncommutativity.  
\item The non planar contribution explodes at small momentum $q_3$, therefore the UV/IR mixing is present.  Indeed, in the commutative case both planar and nonplanar diagrams diverge quadratically. A presence of the noncommutativity allowed us to decrease the UV divergence of the nonplanar graph from a quadratic to a logarithmic. Nevertheless the price for such an improved UV behaviour is a presence of the mentioned IR divergence at small $q_3$ which is absent in the commutative case. 
\item On the one side the nonplanar contribution~\eqref{nonplanans} explodes also for $\theta\to 0$. In addition to that, the amplitude explodes also for any choice of $\theta$ and $q_3$ such that $\theta q_3= k~ 2 \pi$ with $k$ an arbitrary integer.   
The situation is quite typical for the UV/IR mixing: the  $\theta \longrightarrow 0$ limit does not commute with the large $\Lambda$ asymptotics, in particular the latter does not exhibit a smooth commutative limit.

\item{The nonplanar correction to the propagator is not proportional to $\delta(q-s)$, what implies that the ``deformed momentum conservation law" which holds at the classical level is anomalously broken by quantum corrections.}
\end{itemize}

\section{The 3 dimensional case}\label{threed} 
 When the number of the dimensions decreases, the UV properties of QFT usually improve, therefore it is interesting to study the UV/IR issue in the 3-dimensional situation. There is an important point, however.

On the one side in the 2+1 dimensional case there is no third spatial direction $x_3$, therefore  $x_0$ is the most natural candidate to be a noncommutative coordinate in the commutation relations.  On the other side, as we commented in the previous section, the noncommutativity of the time variable $x_0$ leads to difficulties in the \emph{Lorentzian} setting: the Wick rotation, generally speaking, does not help to compute the loop integrals.

Therefore in this section we consider the \emph{Euclidean} 3-dimensional QFT  described by the classical Lagrangian
\bea
S[\phi] = \int_{\mathbb{R}^3} \dd^3x\, \left(\frac{1}{2}\partial_{\mu}\phi(x)\star\partial^{\mu}\phi(x) + \frac{1}{2} m^2 \phi(x)\star\phi(x) + \frac{\lambda}{4!}\phi(x)^{\star 4}\right), \label{action2}
\eea
where the coordinates are $x_1$, $x_2$ and $x_3$.
Note that in such a situation the corresponding deformed symmetry is described by the twisted Poincar\'e Hopf algebra, which we discussed in \ref{TPHa}, however
one has to replace $\eta_{\mu\nu}$ by $\delta_{\mu\nu}$ whenever it appears. 
 In this section the commutation relations will be 
\bea
{}[x^3 \ds x^1]&=&-\ii\theta x^2,\nonumber\\
{}[x^3 \ds x^2]&=&\ii\theta x^1. \label{commrel0}
\eea

We will focus on the nonplanar contribution only, since the planar correction is the same as in the commutative case. The main difference with respect to the 3+1 dimensional case is that the divergences for  this case are linear in $\Lambda$, not quadratic.
The nonplanar contribution can be elaborated along the lines of the previous section,  eliminating $p_3$ and $\int d p_3$, whenever they appear.  Note that no Wick rotation is needed at the intermidiate stage, since the theory is Euclidean from the very beginning.
In particular introducing the cutoff $\Lambda$ in the momenta space one arrives at:
\be
\mbox{NPl.} = \frac{1}{q^2 +m^2}\frac{1}{s^2+m^2} \frac{\pi^2\delta(q_3 - s_3)}{2\, \left( \sin \left( \frac{\theta q_3}{2} \right)  \right) ^{2}} \int_{-\Lambda}^{\Lambda} dp_3 \,\,\frac{1}{p_3^2 + \left(\mathcal{P}_1(q,s,p_3)\right)^2 + \left(\mathcal{P}_2(q,s,p_3)\right)^2 +m^2}.
\ee
The crucial difference with respect to the commutative case is the fact that the integral is UV finite. From now on we consider this contribution at  $\Lambda\longrightarrow\infty$. Unfortunately the combination
$\left(\mathcal{P}_1(q,s,p_3)\right)^2 + \left(\mathcal{P}_2(q,s,p_3)\right)^2$ is too complicated to allow for an exact calculation of the  integral  (c.f. \eqref{p1p2}), however we have the nice estimate \eqref{estimate}, which implies
\bea
{\frac {\pi \, \left| \sin \left(\frac{\theta q_3}{2} \right)
 \right| }{\sqrt {{m}^{2} \left( \sin \left(\frac{\theta q_3}{2} \right)  \right) ^{2}+ \frac{\left( |q|+|s| \right)^{2}}{4}  }}} &\leq& \int_{-\infty}^{\infty} dp_0 \,\,\frac{1}{p_3^2 + \left(\mathcal{P}_1(q,s,p_3)\right)^2 + \left(\mathcal{P}_2(q,s,p_3)\right)^2 +m^2} \nonumber \\
&\leq&{\frac {\pi \, \left| \sin \left(\frac{\theta q_3}{2} \right)
 \right| }{\sqrt {{m}^{2} \left( \sin \left(\frac{\theta q_3}{2} \right)  \right) ^{2}+ \frac{\left( |q|-|s| \right)^{2}}{4}  }}}.
\eea
Note that this nonplanar correction is still singular at
\be
\sin \left(\frac{\theta q_3}{2} \right) \longrightarrow 0,
\ee
in particular the commutative limit of this expression does not exist.

The nonplannar graph of the corresponding commutative theory is linearly divergent in $\Lambda$. The fact that the angular noncommutativity allowed us to trade the UV divergence of the commutative theory for the IR divergence of noncommutative theory is nothing but the UV/IR mixing.

\section{Particle Decay}\label{particle}

In this section we discuss an application of the $\star$-sum of momenta (\ref{StarSumFull}) to the kinematics of particles decays. Before we start, let us point out that the dispersion relation of a particle of mass $m$ is undeformed. This can be traced back to the fact that the free propagator is unchanged in our model:
\be
E^2=|\vec{p}|^2+m^2 .\nn
\ee

Let us assume that a particle of mass $M$ and the momentum $p$ moves along the $3$-axis and it decays into two particles  with momenta $q$ and $r$ and the corresponding masses $m_q$ and $m_r$. Equation \eqref{three} modifies the momentum conservation law for three particles. Note that in \eqref{three} all momenta are incoming. For the case at hand it is natural to change sign of outgoing momenta $q$ and $r$. This leads to
\be
p+_\star (-q)+_\star (-r)=R(-q_3-r_3)p-R(-p_3-r_3) q-R(-p_3+q_3)r =0. \label{ConsLaw}
\ee
In the chosen coordinate system momenta are given by
\be
p=\begin{pmatrix}
\sqrt{M^2+p^2_3}\\ 0\\ 0\\ p_3
\end{pmatrix},\quad
q=\begin{pmatrix}
E_q\\ \vec{q}
\end{pmatrix}.\quad
r=\begin{pmatrix}
E_r\\ \vec{r}
\end{pmatrix} . \label{Momenta}
\ee
From (\ref{ConsLaw}) four equations follow:
\bea
&&\sqrt{M^2+p^2_3}=E_q+E_r ,\\
&&p_3=q_3+r_3 ,\\ 
&&0=\cos{\left(\frac{\theta}{2}(p_3+r_3)\right)}q_1-\sin{\left(\frac{\theta}{2}(p_3+r_3)\right)}q_2+\cos{\left(\frac{\theta r_3}{2}\right)}r_1-\sin{\left(\frac{\theta r_3}{2}\right)}r_2, \label{1}\\
&&0=\cos{\left(\frac{\theta}{2}(p_3+r_3)\right)}q_2+\sin{\left(\frac{\theta}{2}(p_3+r_3)\right)}q_1+\cos{\left(\frac{\theta r_3}{2}\right)}r_2+\sin{\left(\frac{\theta r_3}{2}\right)}r_1 .\label{2}
\eea
Squaring and adding (\ref{1}) and (\ref{2}), we obtain 
\be\label{square}
|{\vec q}_{12}|^2=|\vec{r}_{12}|^2,
\ee
where ${\vec q}_{12}$ is the projection of the spatial part of the momentum $q$ on the $12$-plane and analogously for ${\vec r}_{12}$. Using this result we calculate the components $q_3$ and $r_3$ to be
\bea
&& q_{3}=\frac{p_3(M^2+m_q^2-m_r^2)\mp\sqrt{(p_3^2+M^2)[(M^2+m_q^2-m_r^2)^2-4M^2(m_q^2+r^2_{12})]} }{2M^2},\\
&& r_{3}=\frac{p_3(M^2+m_r^2-m_q^2)\pm\sqrt{(p_3^2+M^2)[(M^2+m_q^2-m_r^2)^2-4M^2(m_q^2+r^2_{12})]}}{2M^2}.
\eea
There is no deformation in these equations, that is the components $q_3$ and $r_3$ are the same as in the commutative case.

Going back to equations (\ref{1}) and (\ref{2}), we observe that they can be written in the matrix form as
\be\label{rot}
0 = \mathcal{R}\left(\frac{\theta p_3}{2}\right)\vec{q}_{12}+\vec{r}_{12},
\ee
where the rotation matrix in $12$-plane is given by
\be\mathcal{R}\left(\frac{\theta p_3}{2}\right)=\left(
\begin{array}{cc}
\cos{\left(\frac{\theta p_3}{2}\right)} & -\sin{\left(\frac{\theta p_3}{2}\right)} \\
\sin{\left(\frac{\theta p_3}{2}\right)}  & \cos{\left(\frac{\theta p_3}{2}\right)}  
\end{array}
\right) .
\ee
From \eqref{rot} it follows that the angle between momenta ${\vec q}_{12}$ and ${\vec r}_{12}$ in $12$-plane will be $\Delta\varphi = \pi-\frac{\theta p_3}{2}$. In the commutative case this angle is $\Delta\varphi = \pi$ since the decaying particle $p$ moves in the $3$-direction and has no momenta in the $12$-plane. The noncommutativity changes this result and produces a non-zero momentum in the $12$-plane after the decay. This correction (the resulting momentum in $12$-plane) depends on $p_3$. Especially, for $\theta p_3 = 2k\cdot 2\pi$ there will be no correction at all, that is particles $q$ and $r$ will move back to back after the decay. On the other hand, if $\theta p_3 = (2k+1)\cdot 2\pi$ the correction is maximal, that is particles $q$ and $r$ move in the same direction in $12$-plane.

We also observe that if a particle $p$ decays from rest, $p_3=0$, there will no correction to the commutative result. On the other hand, if we boost to a reference frame which moves in the $3$-direction, we observe the non-zero correction given by (\ref{rot}). This suggests a non-trivial connection between the boost $M_{03}$ and the rotation in the $12$-plane $M_{12}$. From the coproduct (\ref{TwistedCoproductLorGen}) it is obvious that these two generators are entangled. A better understanding of this entanglement and its relation to the result (\ref{rot}) is needed and we plan to address this problem in our future work.

Finally, let us comment these results in the $x^0$ variant of (\ref{angtwist}). In this case, the equation corresponding to (\ref{rot}) is given by
\be\label{rotx0}
0 = \mathcal{R}\left(\frac{\theta E_p}{2}\right)\vec{q}_{12}+\vec{r}_{12},
\ee
where the rotation matrix in $12$-plane is given by
\be\mathcal{R}\left(\frac{\theta E_p}{2}\right)=\left(
\begin{array}{cc}
\cos{\left(\frac{\theta E_p}{2}\right)} & -\sin{\left(\frac{\theta E_p}{2}\right)} \\
\sin{\left(\frac{\theta E_p}{2}\right)}  & \cos{\left(\frac{\theta E_p}{2}\right)}  
\end{array}
\right) .
\ee
The corresponding angle between momenta ${\vec q}_{12}$ and ${\vec r}_{12}$ in $12$-plane is $\Delta\varphi = \pi-\frac{\theta E_p}{2}$, that is, it depends on the energy $E_p$ of the decaying particle. It is interesting to note that in this case, a non-zero noncommutative correction exists also in the reference frame in which the decaying particle is at rest and it is given by $\Delta\varphi = \pi-\frac{\theta M}{2}$.

\section{Conclusions}
We studied a scalar field theory in the presence of a noncommutative product of the angular kind in three and four dimensions.
The model is twist-Poincar\'e invariant and exhibits UV/IR mixing with analogies with the translation invariant cases, in particular again the planar diagram is unchanged with respect to the commutative case, the novelties manifesting themselves for the nonplanar case.  

We also studied the decay of particles in this noncommutative space, considering the two cases time commutative or not. The theory breaks rotational invariance, replacing it with a deformed Hopf algebra, as a consequence momentum is not conserved and decays of particles at rest (in the noncommutating time case) are not ``back to back", although there is not a preferred direction of decay. For the case of spatial noncommutativity there is no deformation for particles at rest, but there is for particle in motion, which means that the action of boosts, which is deformed, is nontrivial, and this issue deserves further scrutiny.

Another point which may give interesting developments is the following. Angular noncommutativity could be obtained as a suitable limit of a fully noncommuting algebra, $\R^3_\lambda$, based on $\mathfrak{su}(2)$. Corresponding field theories on $\R^3_\lambda$ do not exhibit mixing. It would be therefore interesting to regard the field theory analyzed in the present paper through the limiting procedure we have just mentioned.

\appendix

\section{Product of plane waves}
In our case we want to calculate
\be
\e^{\ii p \cdot x}\star\e^{\ii q \cdot y}=\exp{\left\{\frac{\theta q_3}{2}\left(x^2\partial_{x^1} - x^1 \partial_{x^2}\right)
\right\}}e^{\ii p \cdot x}
\exp{\left\{\frac{\theta p_3}{2}\left(y^2\partial_{y^1} - y^1 \partial_{y^2}\right)
\right\}}e^{\ii q \cdot y}    \label{interm1}
\ee

In order to find the relation for our $\star$ product we need two ``tricks''.
\subsubsection*{Trick \# 1.}
If the three operators $A$, $B$ and $C$ form the Lie algebra:
\be
[A,B] = C, \quad [A,C] = - \lambda A, \quad [B,C] = \lambda B, \label{algebra}
\ee 
where $\lambda$ is some C-number, then for an arbitrary C-number $\tau$ the following identity holds:
\be
 \exp{\left\{\tau\left(A+B\right)\right\}} =  \exp{\left\{ \alpha(\tau) B\right\}}\, \exp{\left\{ \beta(\tau) C\right\}}\, \exp{\left\{ \gamma(\tau) A\right\}}, \label{trick1}
\ee
where
\be
\alpha(\tau) = \gamma(\tau) = \sqrt{\frac{2}{\lambda}} \,\tanh{\left(\sqrt{\frac{\lambda}{2}}\cdot\tau\right)},\quad
\beta(\tau) = \frac{2}{\lambda} \,\ln \cosh \left(\sqrt{\frac{\lambda}{2}}\cdot\tau\right)
\ee
Proof: See the appendix of the book \cite{Kirzhnits}.
\subsubsection*{Trick \# 2.}
For an arbitrary function $\mathcal{F}(z)$ the following relation takes place:
\be
\exp{\left\{k z\partial_z\right\}}\,\, \mathcal{F}(z) = \mathcal{F}(\exp(k)\,z) \label{trick2}
\ee
Proof:\footnote{This proof can be found  in \cite{Kirzhnits}.} Let us introduce the new variable $w= \ln(z)$ and use the fact that derivatives generate translations.

Now we are ready for the next steps. The first factor of the righthand side of~\eqref{interm1} has exactly the structure of the lefthand side of \eqref{trick1} for
\be
A = x^2\partial_{x^1}, \quad B = -x^1\partial_{x^2}, \quad \tau = \frac{\theta q_3}{2}.
\ee
Computing the commutator of $a$ and $b$ one immediately arrives to the Lie algebraic structure \eqref{algebra} at
\be
C  = x^1\partial_{x^1} - x^2\partial_{x^2}, \quad \lambda = -2.
\ee
Again~\eqref{trick1} tells us that 
\bea
\exp{\left\{\frac{\theta q_3}{2}\left(x^2\partial_{x^1} - x^1 \partial_{x^2}\right)\right\}}  =
(\alpha-\mbox{term})\,(\beta-\mbox{term})\,(\gamma-\mbox{term})
\eea
where
\bea
 (\alpha-\mbox{term})&\equiv& \exp{\left\{ \tan\left(\frac{\theta q_3}{2}\right) x^1\partial_{x^2}\right\}}, \nonumber\\
  (\beta-\mbox{term})&\equiv & \exp{\left\{ -\ln{\left(\cos{\left(\frac{\theta q_3}{2}\right)}\left(x^1\partial_{x^1} - x^2\partial_{x^2}\right)\right)}\right\}}, \nonumber\\
   (\gamma-\mbox{term})&\equiv& \exp{\left\{ -\tan\left(\frac{\theta q_3}{2}\right) x^2\partial_{x^1}\right\}}. \label{abgterms}
\eea
The three operators defined by \eqref{abgterms} act on  plane waves in the following way:
\be
(\gamma-\mbox{term})e^{-\ii px}   =   e^{-\ii p' x}, \quad p' = \left(\begin{array}{c} p_0 \\ p_1 \\ p_2 - \tan{\left(\frac{\theta q_3}{2}\right)} p_1 \\ p_3  \end{array} \right) .\label{step1}
\ee
Using the second trick \eqref{trick2} we immediately find that
\be
(\beta-\mbox{term})e^{-\ii p'x}   =   e^{-\ii p'' x}, \quad p'' = \left(\begin{array}{c} p'_0 \\ 
\left(\cos{\left(\frac{\theta q_3}{2}\right)}\right)^{-1}p'_1 \\ \cos{\left(\frac{\theta q_3}{2}\right)}p'_2  \\ p'_3  \end{array} \right) . \label{step2}
\ee
Finally
\be
(\alpha-\mbox{term})e^{-\ii p''x}   =   e^{-\ii p''' x}, \quad p''' = \left(\begin{array}{c} p''_0 \\ p''_1 +\tan{\left(\frac{\theta q_3}{2}\right)} p''_2 \\ p''_2  \\ p''_3  \end{array} \right) .\label{step3}
\ee
Combining \eqref{step1}, \eqref{step2} and \eqref{step3} together we arrive to \emph{deformed}  sum:
\be
e^{-p \cdot x} \star e^{-\ii q\cdot  x} = e^{-(p+_{\star}q)\cdot  x}, 
\ee
where the $\star$-sum of the 4-momenta  is defined as follows:
\be
p+_{\star}q = R(q_3)p + R(-p_3)q, 
\ee
and
\be
R(t) \equiv \left(
\begin{array}{cccc}
1 & 0 & 0 & 0 \\
0 & \cos{\left(\frac{\theta t}{2}\right)} & \sin{\left(\frac{\theta t}{2}\right)} & 0 \\
0 & -\sin{\left(\frac{\theta t}{2}\right)}  & \cos{\left(\frac{\theta t}{2}\right)} & 0 \\
0 & 0 & 0 & 1
\end{array}
\right) .
\ee


\subsection*{Acknowledgments}
The authors acknowledge the COST action QSPACE. MK, FL and PV acknowledge the support of  the INFN Iniziativa Specifica GeoSymQFT; FL the Spanish MINECO under project MDM-2014-0369 of ICCUB (Unidad de Excelencia `Maria de Maeztu'). The work of MDC and NK is supported by project
ON171031 of the Serbian Ministry of Education and Science. MDC would like to thank the University Federico II for kind hospitality supported by a \emph{visita di breve mobilit\`a} and NK acknowledges the support from ERASMUS+ KA1 programme.

\end{document}